\documentstyle[11pt,newpasp,twoside,epsf]{article}
\def\edcomment#1{\iffalse\marginpar{\raggedright\sl#1\/}\else\relax\fi}
\reversemarginpar

\begin{document}
\title{Blobby accretion in magnetic cataclysmic variables}
 \author{A.V. Halevin, I.L. Andronov}
\affil{Department of Astronomy, Odessa National University,
T.G.Shevchenko park, 65014, Odessa, Ukraine,
halevin@astronomy.org.ua}
 \author{N.M. Shakhovskoy}
\affil{Crimean Astrophysical Observatory, Nauchny 98409 Crimea,
Ukraine}
 \author{S.V. Kolesnikov, N.I. Ostrova}
\affil{Astronomical Observatory, Odessa National University,
T.G.Shevchenko park, 65014, Odessa, Ukraine}

\begin{abstract}
The processes of accretion of the gaseous blobs with different
masses and densities onto strongly magnetized white dwarfs in the
systems of polars have been modeled. We have proved that shot
noise in blue wavelengths represents accretion of the smaller and
denser blobs than in redder wavelengths. Using combined "smooth
particle hydrodynamics - drag force" model, we have predicted a shape of
the accretion stream and active regions on the white dwarf
surface.
\end{abstract}

\section{Introduction}

The idea about the blobby accretion in magnetic cataclysmic
variables had appeared as an attempt to explain the flickering and
the soft X-ray excess in such systems. The fast variability at a
time-scale of dozens of seconds is well described by the ``shot
noise'' model with an exponential decay of its auto-correlation
function (ACF, Andronov 1994). The ``shot noise'' is interpreted
as a result of accretion of large diamagnetic blobs (Beardmore \&
Osborne 1997; Kuijpers \& Pringle 1982; Panek 1980).

Such blobs originate due to the Rayleigh-Taylor instabilities,
when a heavy fluid (matter) is opposed in the gravitation field to
a light fluid (magnetic field).

The dynamical properties of such blobs were investigated by King
(1993). The drag force, which has an influence on the trajectories
of the blobs in magnetic field is expressed as follows:
\begin{equation}
f_{drag}=-k[\vec \nu' - (\vec \nu' \cdot \vec b')\vec b']
\end{equation}

The drag coefficient
\begin{equation}
k = \frac{B^{2}l^{2}}{c_{A}m}
\end{equation}
is dependent on such parameters, as the magnetic field strength
$B$, blob size $l$, blob mass $m$ and Alfv\'en velocity in interblob
plasma $c_{A}$. Because the Alfv\'en velocity and the mass of the
blob are expressed as
\begin{equation}
c_{A} = \frac{B}{\sqrt{4\pi \rho_{i}}}, ~~~m = \frac{4}{3}\pi
l^{3} \rho_{b}
\end{equation}
where $\rho_{b}$ and $\rho_{i}$ are the densities of the blob and
interblob plasma in the flow, respectively, the drag coefficient is
expressed as
\begin{equation}
k = \frac{3B\rho^{1/2}}{2\pi^{1/2}\rho_{b}l}
\end{equation}
In the models of accretion, the dipolar field
configuration is usually assumed, although, it is not a precise approach.

Varying the parameters of density and size, we can achieve a good
approach to the observations, comparing blob velocities with
existing Doppler tomograms (e.g. Heerlein et al. 1999).

Possible differences between the blob parameters, such as mass and
size (Wynn \& King 1995), could lead the differences in locations
of the active regions on the white dwarf surface and in the variability
of the shot noise decay time.

If we calculate the blob trajectories with the range of sizes
between $10^{8}$ to $10^{10}$ cm, then in the case of HU~Aqr, we
can obtain the next ideal dependence of the initial blob size on
the azimuth on white dwarf surface (Fig.1).

\begin{figure}
\plotone{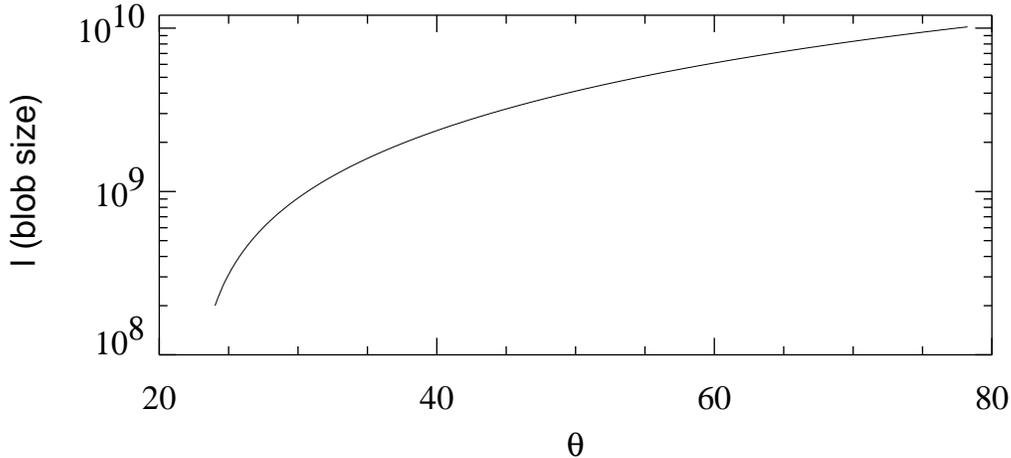} \caption{Dependence of the initial size of the
accreted blobs on the longitude at the white dwarf surface.}
\end{figure}

Having such a clear picture of distribution of the blob sizes at the
white dwarf surface, we can expect the smooth variability of the
shot noise decay time. If we remove the orbital variability from
our observations, we can calculate the biased ACF and, using the
method described by Andronov (1994), we can calculate the
exponential decay time for unbiased ACF.

During the fall onto the white dwarf's surface, the blobs are
deformed owed to tidal forces. The undisturbed size of a blob is
expressed by Halevin et al.(2002b) as
\begin{equation}
l = \frac{\tau}{4}\left(\frac{2GM_{wd}}{r_{coup}}\right)^{1/2}
\end{equation}
where $r_{coup}$ is the coupling distance and $M_{wd}$ is a mass
of the white dwarf. One can see that the falling time of the blobs
does not depend on the stopping height.

\begin{figure}
\plotone{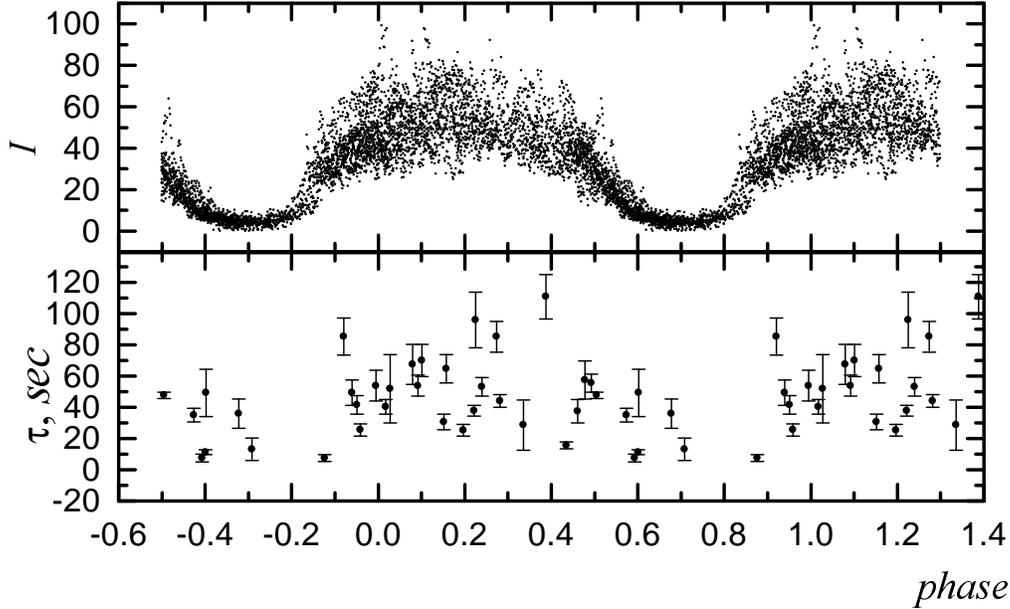} \caption{Phase curves for count rate and $\tau$
for Ginga satellite observations of AM~Her. Period is 0.128927041
days (Greeley et al. 1999).}
\end{figure}
\begin{figure} \plotone{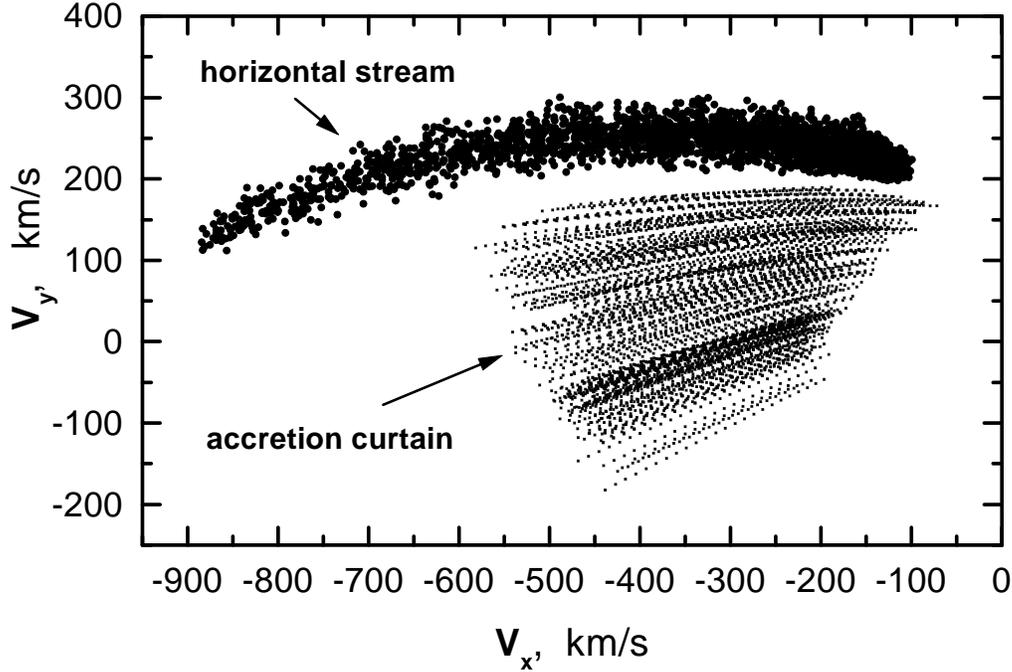}
\caption{Positions of the SPH points in velocity space for
HU~Aqr.}
\end{figure}

In the Fig.2, one can see the phase variability of the shot noise decay
time for X-ray Ginga observations of AM~Her. Although the data are
significant, we do not see smooth variations, as expected.

To make more realistic model of blobby accretion in polars, we have
used the SPH method. The pure hydrodynamical model was made by
Cach \& Howell (2002). Dynamical positions of the flow points at the
Doppler tomogram is shown in Fig.3.

In our model, we have used the hydrodynamical parameters of the
flow to make comparison between the gas pressure (not the ram one)
and the magnetic pressure. When the magnetic pressure becomes
larger than the gas pressure, we convert our SPH points
into blobs.

Here we used computed from our model parameters of the flow with
the formula of Hameury, King, \& Lasota (1986) for the minimum
scale of the structures which are unstable to the Rayleigh-Taylor
mechanism
\begin{equation}
l = \frac{B^{2}_{b}r^{2}_{b}}{2GM_{wd}\rho_{b}}
\end{equation}
where $B_{b}$ is magnetic field and $r_{b}$ is the distance of
blob appearance. Substituting (6) to the (4), we obtain the
expression for the drag force in the next form:
\begin{equation}
k = \frac{4\pi^{1/2}GM_{wd}B\rho^{1/2}_{i}}{r^{2}_{b}B^{2}_{b}}
\end{equation}
The very interesting result is that, under such assumptions, the
drag force does not depend neither on the blob sizes nor on the
blob density, but only on the magnetic field and the interblob plasma
density, which we assumed to be a constant.

The modeled Doppler tomogram for HU~Aqr is shown in Fig.4.

\begin{figure}
\plotfiddle{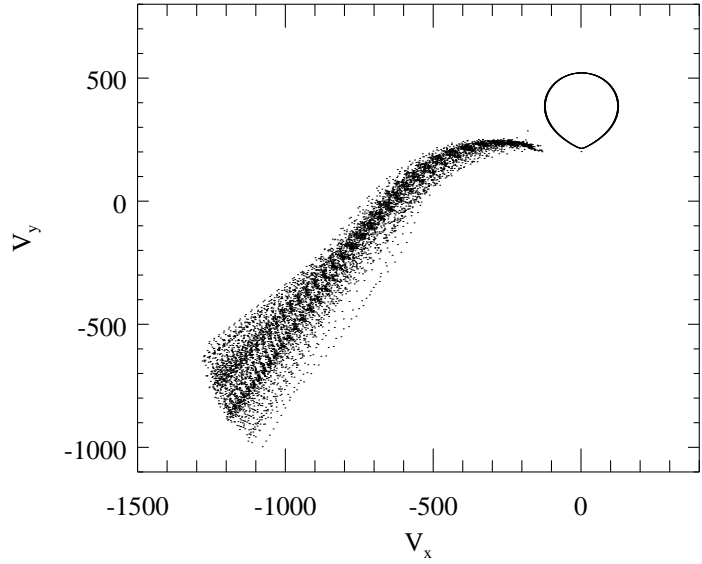}{2.7in}{0}{70}{70}{-160}{-250}
 \caption{Doppler tomogram for combined SPH-"drag force"
model in the case of HU~Aqr.}
\end{figure}

First of all, one can see that this model cannot explain the
observed accretion curtain features. It will be possible, if we
take into account the disruption of the blobs by the Kelvin-Helmholtz
instability. In the next Fig.5, one can see the distribution of the
accretion blob sizes on the white dwarf surface for the case of
AM~Her. There is already no strong dependence. However, possible
effect is that large blobs can originate in outer parts of the
accretion stream, and, hence, they will fall into outer parts of the
active region on white dwarf surface.

\begin{figure}
\vspace{0.5in} 
\plotfiddle{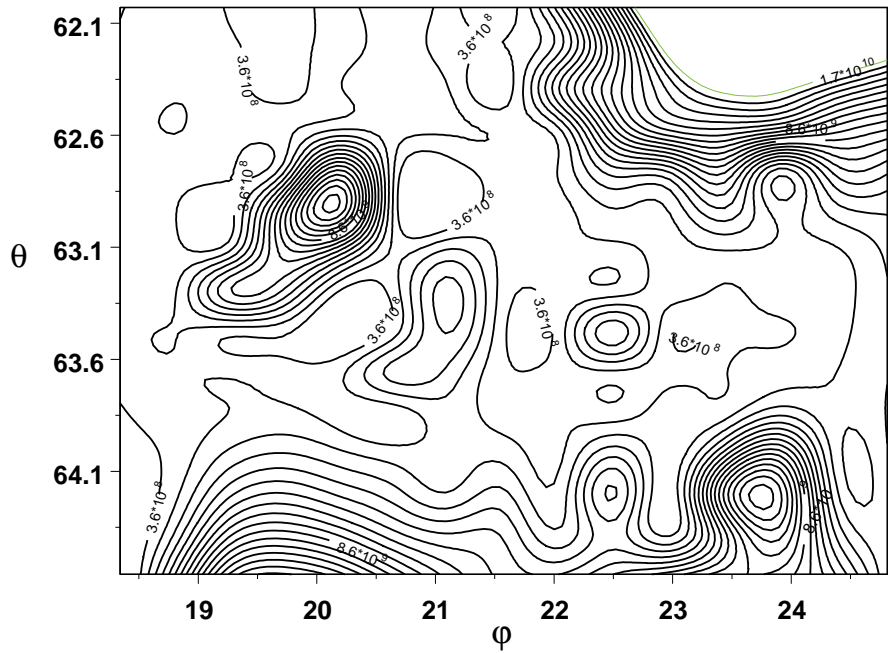}{2.7in}{0}{70}{70}{-190}{-50}
\caption{Initial size (cm) distribution of the accreted blobs on
the white dwarf surface in the case of AM~Her.}
\end{figure}

Let analyze again the decay scale variability curves. The composite
curves do not show strong dependencies. But some individual curves
(as in the case of BY~Cam) can show them.

Much more simpler can be the picture for the soft X-ray shot noise,
because the source of this radiation is expected have a zero (or
about) height.

Take a look at the size distribution (Fig.5). Because the sizes of
the blobs are in inverse proportion with the blob density and the
density is inversely proportional to the shock height, we can
expect that this distribution is about to express also the shock
height. Such distribution of the shock heights could explain the
observed double humped orbital soft X-ray variability of AM~Her
(Christian 2000) as absorption by higher shocks.

Under our assumptions, larger blobs have smaller densities. As we
know from the work of Fisher \& Beuermann (2001), for larger
accretion rate per unit area, the maximum of cyclotron radiation
shifts to the shorter wavelengths. So we can expect, that in blue
wavelengths we can observe accretion of the shorter blobs, that has been
already discovered by Beardmore \& Osborne (1997).

\begin{figure}[t]
\plotfiddle{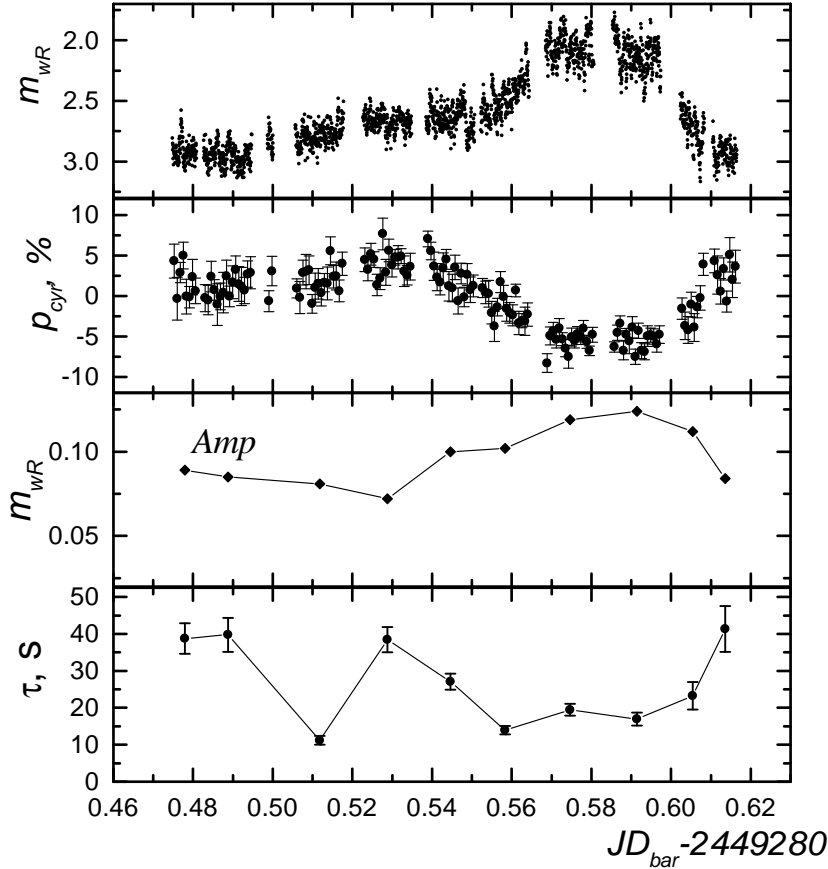}{4.3in}{0}{60}{60}{-190}{-130} \caption{Phase
curves for a magnitude, polarization, shot noise amplitude and
$\tau$ for 2.5-m telescope observations of the BY~Cam (Halevin et
al. 2002a).}
\end{figure}

So, let me show the final results of investigations of the shot
noise decay time in magnetic cataclysmic variables (Table 1). For
AM~Her and EF~Eri, we have analysed the X-ray variability and
detected the mean decay times of about 70 s.

\begin{table}
\caption{Mean decay time estimates.} \label{tab1}
\begin{tabular}{lllcr}
\hline Star&$\tau$, s&wavelengths&Magn. field, G&ref.$^{*}$\\
\hline AM~Her&67.5$\pm$9.2&X-ray,~1.7-10.4~keV&13&H1\\
AM~Her&70&optical~I,R~bands&13&BO\\
AM~Her&25&optical~U~band&13&BO\\
EF~Eri&69.0$\pm$11.0&X-ray,~1.7-10.4~keV&16&H1\\
BY~Cam&40.8$\pm$5.1&optical~V,R~bands&40&H1\\
QQ~Vul&42.8$\pm$5.7&optical~V,R~bands&35&H2\\
AR~UMa&no~shot~noise?&optical~V,R~bands,~low~state&250&SH\\ \hline
\end{tabular}
$^{*}$ H1 - Halevin et al. (2002a), H2 - Halevin et al. (2002b),
BO - Beardmore \& Osborne (1997), SH - Shakhovskoy \& Halevin
(2000).
\end{table}

From optical observations in V and R band for BY~Cam and QQ~Vul,
the systems with higher magnetic fields, the decay time is smaller.
For AR~UMa, we could not find the shot noise because the system
was in its low state. But it can be the consequence of the high
magnetic field and the full disruption of the blobs by the
Kelvin-Helmholtz instability.

So, the main our future perspectives are:\\ $\bullet$Advanced
models with Kelvin-Helmholtz disruption.\\ $\bullet$Investigations
of the shot noise behavior in soft X-rays.\\
$\bullet$Investigations of the systems with different magnetic
fields, especially high-field polars.\\ $\bullet$Analysis of the
long term behavior of the shot noise parameters.

\end{document}